# Coronal Fe XIV Emission During the Whole Heliosphere Interval Campaign

Richard C. Altrock
Air Force Research Laboratory
Space Weather Center of Excellence
PO Box 62, Sunspot, NM 88349, USA

**Abstract.** Solar Cycle 24 is having a historically long and weak start. Observations of the Fe XIV corona from the Sacramento Peak site of the National Solar Observatory show an abnormal pattern of emission compared to observations of Cycles 21, 22, and 23 from the same instrument. The previous three cycles have shown a strong, rapid "Rush to the Poles" (previously observed in polar crown prominences and earlier coronal observations) in the parameter N($t,l,dt$) (average number of Fe XIV emission features per day over $dt$ days at time $t$ and latitude $l$). Cycle 24 displays a weak, intermittent, and slow "Rush" that is apparent only in the northern hemisphere. If the northern Rush persists at its current rate, evidence from the Rushes in previous cycles indicates that solar maximum will occur in early 2013 or late 2012, at least in the northern hemisphere. At lower latitudes, solar maximum previously occurred when the time maximum of N($t,l,$365) reached approximately 20° latitude. Currently, this parameter is at or below 30°and decreasing in latitude. Unfortunately, it is difficult at this time to calculate the rate of decrease in N($t,l,$365). However, the southern hemisphere could reach 20° in 2011. Nonetheless, considering the levels of activity so far, there is a possibility that the maximum could be indiscernible.

## 1. Introduction

Trellis (1954, 1963), Leroy and Noens (1983), and Wilson *et al.* (1988) (and references therein) discussed the high-latitude "extended" solar cycle seen in the Fe XIV corona prior to the appearance of sunspots and active regions at lower latitudes. Altrock (1997) described persistent coronal emission features that appeared near 70° latitude in 1979 and 1989 and slowly migrated toward the equator, merging with the latitudes of sunspots and active regions after several years. Wilson *et al.* (1988) discussed other observational parameters that have similar properties, and this information was updated by Altrock, Howe, and Ulrich (2008) for torsional oscillations.

Altrock (2007) showed that the high-latitude emission features are situated above the high-latitude neutral line of the large-scale photospheric magnetic field seen in Wilcox Solar Observatory synoptic maps, thus implying a connection with the solar dynamo.

Altrock (2003) discussed coronal emission features seen in Fe XIV which, prior to solar maximum in Cycles 21 - 23, appeared above 50° latitude and began to move toward the poles at a rate of 8° to 11° year$^{-1}$. This motion continued for three or four years, at which time the emission features disappeared near the poles. This phenomenon is called the "Rush to the Poles" (Rush), and it was first identified in polar crown prominences and first observed in the corona by Waldmeier (1957, 1964) (see also Trellis, 1963). Altrock (2003) concluded that *i*) the maximum of solar activity, as defined by the smoothed sunspot number, occurred 1.5 ± 0.2 year before the extrapolated linear fit to the Rush reached the poles, and *ii*) the Rush could be used to predict the date of solar cycle maximum up to three years prior to its occurrence. He stated that, "For solar Cycle 24, a prediction of the date of solar maximum can be made when the Rush becomes apparent, approximately eleven years after its Cycle-23 onset on 1997.58; i.e., in 2008 - 2009. When that occurs, the average slope for Cycles 21 - 23, 9.38 ± 1.71 ° year$^{-1}$, can be used to predict the arrival date of the Rush at the poles, and then the average lag [time between solar maximum and the date the extrapolated linear fit to the Rush reached the poles], 1.52 ± 0.20 year, can be used to predict the date of solar maximum ..." (but see Subsection 4.1).

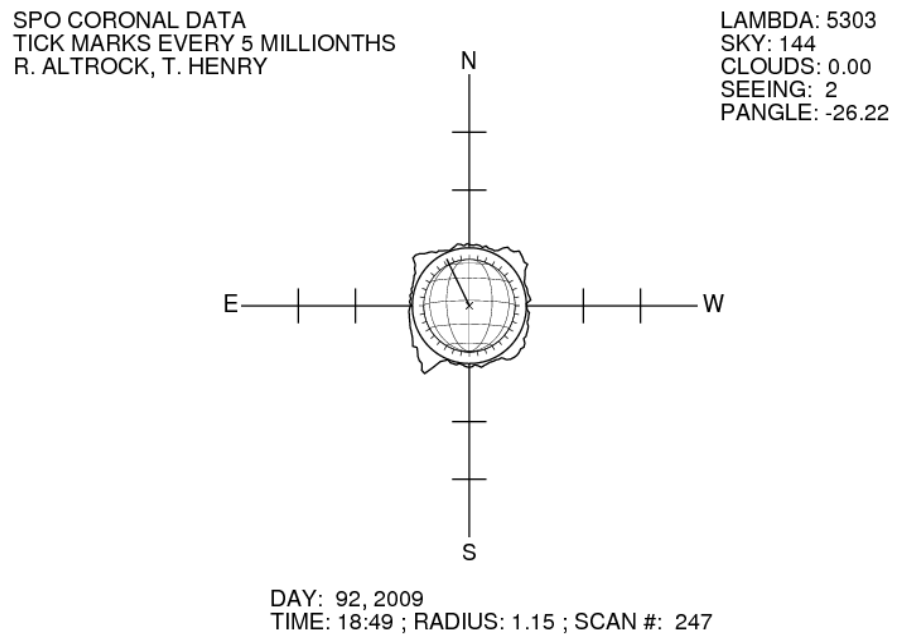

**Figure 1**. Sample polar plot of Fe XIV intensity at 1.15 *Ro* at solar minimum. Intensity is zero at outer circle.

**2. Observations**

Observations of the Fe XIV 530.3 nm solar corona have been attempted three to seven times a week since 1973 with the Photoelectric Coronal Photometer and 40 cm coronagraph at the John W. Evans Solar Facility of the National Solar Observatory at Sacramento Peak (Fisher, 1973, 1974; Smartt, 1982). The photometer automatically removes the highly variable sky background. Scans at 0.15 solar radii [*Ro*] above the limb every 3° in position angle show coronal features overlying active regions, prominences, large-scale magnetic field boundaries, *etc*. Observations near solar minimum continue to show coronal emission features overlying high-latitude neutral lines, even when there are no active regions at the limb. Figure 1 shows a sample solar minimum scan. Note *i*) the lack of low-latitude active region emission features and *ii*) the four emission features occurring at higher latitudes.

## 3. Procedure

As discussed in Altrock (1997), the daily scans of the corona in Fe XIV at 1.15 *Ro* are examined to determine the location in latitude of local intensity maxima such as the four in Figure 1, and the location of each maximum is plotted on a synoptic map of latitude *vs*. time. Figure 2 shows such a synoptic map from 1973 to 2010. Each point represents a single intensity maximum, and these are hereafter referred to as emission features. Note that nowhere in this analysis is the value of the intensity used, and that allows tracking of very faint features. Cycles 21 - 23 are visible, as are the higher-latitude Rushes.

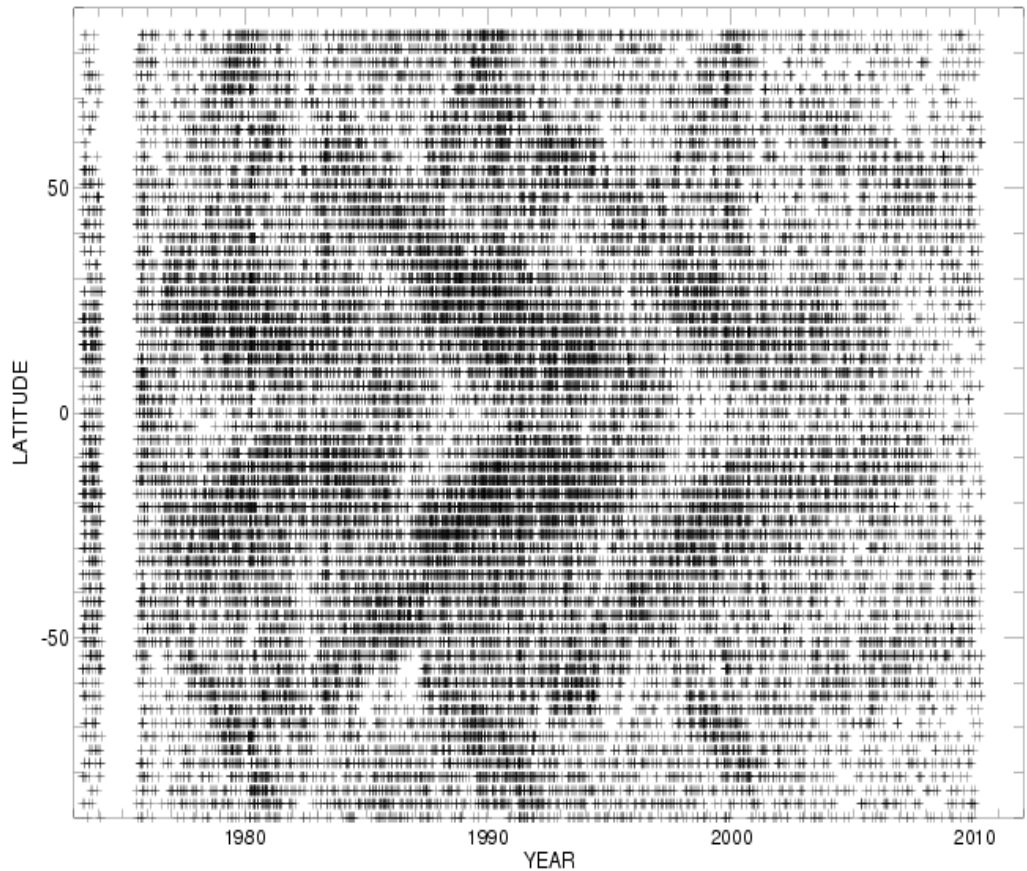

**Figure 2**. Location of 47,000 Fe XIV intensity maxima obtained from scans as in Figure 1 from 1973 - 2010. Note diagonal features running to high and low latitudes.

To clarify the solar cycle behavior of the emission features, we average the number of points at each latitude in Figure 2 over a given time interval. This process allows us to correct the figure for days of missing data, which is an essential step for correctly interpreting the data. Figure 3 shows annual averages of the number of emission features, also averaged over the northern and southern hemispheres.

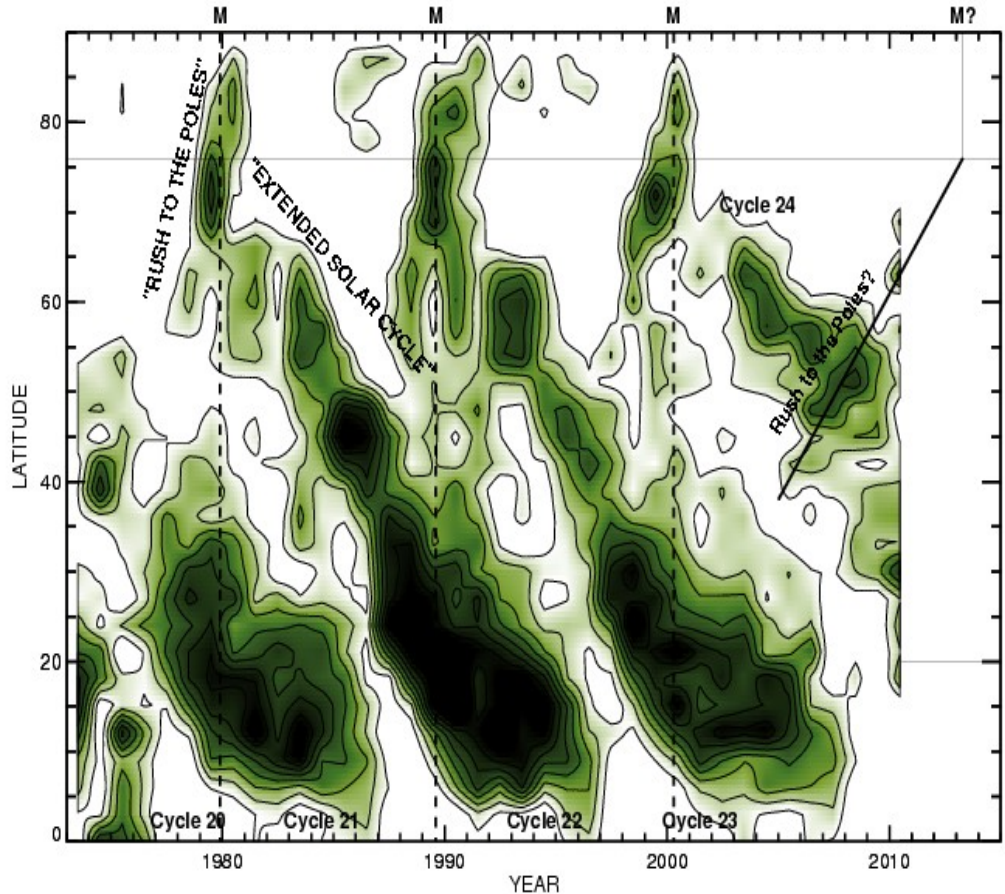

**Figure 3**. Annual northern-plus-southern-hemisphere averages of the number of Fe XIV emission features from 1973 to 2010. The "Rush to the Poles" beginning near 1978 and perhaps near 2005, the extended solar cycle beginning near 1980, and extended solar Cycle 24 beginning near 2000 are indicated. Vertical dashed lines indicate the time of solar maxima. See text for description of other features.

**4. Discussion**

In Figure 3, we can clearly see the nature of extended solar cycles and Rushes over the last 35+ years. Extended solar cycles begin near 70° latitude and end near the equator about 18 years later, as can be seen in Cycles 22 and 23. Note that Cycle 24 began similarly to Cycles 22 and 23; however, its initial rate of migration towards the equator was 40% slower than the previous cycles. The initial rates for Cycles 22 - 24 were -5.3, -4.7, and -3.1 ° year$^{-1}$, respectively. Most recently, emission took a sudden jump down to approximately 30° latitude, and there is a suggestion of a developing Rush. Let us examine this development with a higher-resolution (if noisier) graphic.

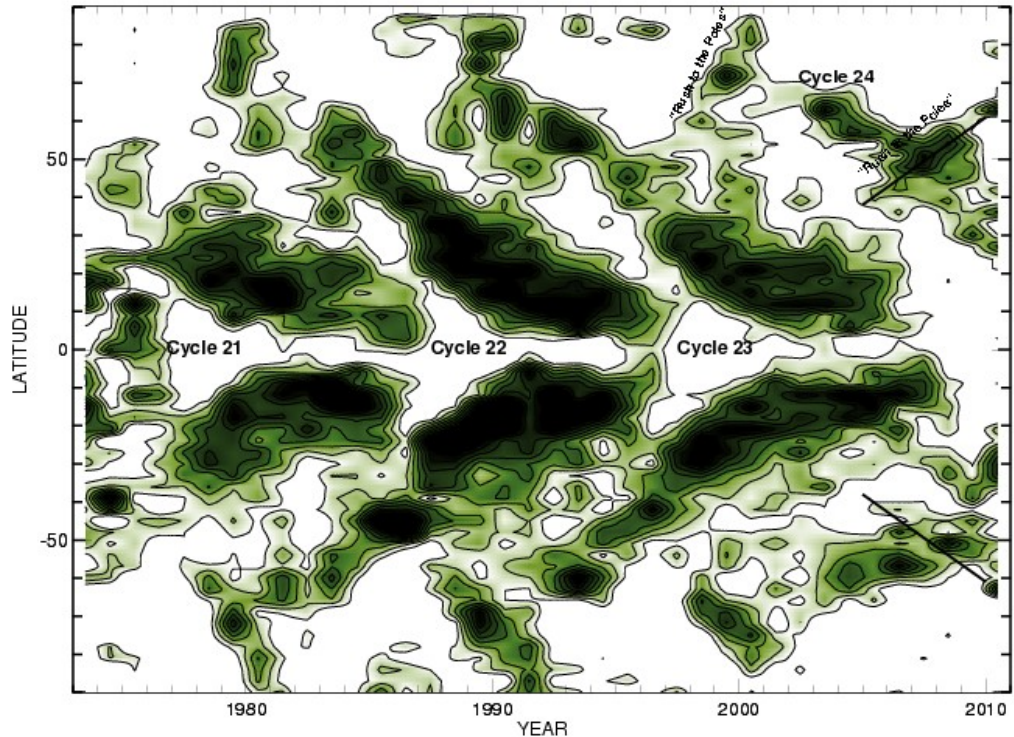

**Figure 4**. Annual averages of the number of Fe XIV emission features from 1973 to 2010 plotted separately for each hemisphere (-90° to +90° latitude). Note the Cycle 24 "Rush to the Poles" indicated by the label in the upper right corner.

4.1. The "Rush to the Poles"

Figure 4 shows the data plotted separately for each hemisphere. We can see the possible appearance in 2005 of a weak Rush in the northern hemisphere, marked by a label, "Rush to the Poles", and the linear fit from Figure 3 in the upper right-hand corner of the graph. The same fit is also plotted in the southern hemisphere, but no equivalent Rush can be clearly distinguished there. Perhaps the current weak Rush is not visible in the southern hemisphere in these noisier data.

In order to address the naturally arising question, "Is this really the Cycle-24 Rush?", we can look at the properties of previous Rushes. In Figure 7 of Altrock (2003), we see the extrapolated Rushes in the previous three cycles (shown by linear fits) from 51° to 90° (the upper half of Figure 3 is similar to that figure, but without the linear fits). The time interval between the extrapolations at 51° in the previous three cycles was 10.8 and 9.6 years. So we might expect the proposed Cycle 24 extrapolated Rush to cross 51° in an approximately similar interval relative to Cycle 23. The proposed Cycle-24 Rush extrapolation crossed 51° in 2007.8, 10.6 year after the Cycle 23 51° crossing. There is no indication in Figures 3 or 4 of a Rush starting at any earlier or later time. This implies that either the proposed Cycle-24 Rush is the actual Cycle 24 Rush, or there is no Cycle 24 Rush (which could imply there will be no Cycle 24 maximum). The lack of a Rush in the southern hemisphere could indicate there will be no measurable maximum there.

With this in mind, we can use the northern hemisphere data as an indicator of when solar maximum will occur, perhaps only in the northern hemisphere.

The estimated current Rush rate is 4.6° year$^{-1}$ (recall 9.4 ± 1.7 ° year$^{-1}$, the average rate in the previous three cycles [Altrock, 2003]). This 50% lower rate makes the earlier suggestion to use the previous higher rate or the previous lag to estimate the time of cycle maximum invalid (see discussion in Introduction).

A more reliable method is to use the property that solar maximum occurs when the linear fit to the Rush reaches a critical latitude. In the previous three cycles, this latitude was 76°, 74°, and 78°, for an average of 76° ± 2° (this can be determined from the figures in Altrock [2003]). Latitude 76° is shown on Figure 3 by a thin line. There, one can see that solar maximum (vertical dashed lines) occurs when the Rushes are near 76°. Figure 3 demonstrates that, at the current rate, the proposed Rush will reach 76° at 2013.3 ± 0.5.

This time can be considered a predicted time of solar maximum in the northern hemisphere.

### 4.2.   Low-latitude emission

The initial rate of migration towards the equator of emission features appears to have been suddenly interrupted around 2009. The higher-latitude emission may be ending, and a new lower-latitude band has developed. This low-latitude band may be able to be used to infer when solar maximum will occur.

In the previous three cycles, solar maximum occurred when the greatest number of Fe XIV emission features, averaged over 365 days and both hemispheres (as in Figure 3), were at latitudes 18°, 21°, and 21° for Cycles 21, 22, and 23, respectively. The average of these values is 20° ± 1.7° (this is somewhat higher than the value of 14°, the mean latitude of sunspot area at solar maximum). Currently, the greatest number of Fe XIV emission features as seen in Figures 3 and 4 is at or below 30° latitude and migrating towards the equator. Unfortunately, there is no clear indication of when the greatest number of Fe XIV emission features will reach 20°. In the southern hemisphere this could be as early as 2011, if the rapid rate seen now in Figure 4 were to continue. So it will be interesting to see if the results from previous cycles hold true for this cycle.

Nothing in this analysis yields the sunspot number to be expected at solar maximum.

Rušin, Minarovjech, and Saniga (2009) using similar data but a different technique predicted two solar maxima: one in the time frame 2010 – 2011 and one in 2012.

## 5. Simulated Coronal Emission

Robbrecht *et al.* (2010) computed simulated Fe XIV emission from magnetic field data. They concluded that high-latitude emission evolves in a U-shaped pattern in latitude vs. time and is unconnected with lower-latitude emission. This implies that there is no extended solar cycle. However, inspection of Figures 2, 3, and 4 here shows that actual measured Fe XIV emission clearly migrates continuously from high latitudes into latitudes where sunspots begin to be seen, and onwards to near the equator.

They also are unable to satisfactorily account for the evolution of torsional oscillations (TO) from near 60° latitude to the equator and activity connected with the TO. Altrock, Howe, and Ulrich (2008) demonstrate that the emission patterns seen in Figure 3 overlie the regions of extreme negative latitude derivative ("shear") of TO from high latitudes down to near the equator and also up into the regions where the "Rush to the Poles" occurs.

It is clear that, when discussing the behavior of Fe XIV emission patterns, it is preferable to use observed data rather than data simulated from magnetic field data.

## 6. Conclusions

The location of Fe XIV emission features in time-latitude space displays an 18-year progression from near 70° to the equator, which has been referred to as the "extended" solar cycle. Cycle 24 emission features began migrating towards the equator similarly to previous cycles, although at a 40% slower rate. In addition, in 2009 the northern hemisphere "Rush to the Poles" became evident and is proceeding at a 50% slower rate than in recent cycles.

Analysis of the northern hemisphere "Rush to the Poles" indicates that solar maximum will occur in early 2013 or late 2012. There is at this time no confirmation of this prediction from the southern hemisphere. This suggests the possibility that there may be no measurable solar maximum in the southern hemisphere.

Low-latitude emission features are migrating towards the equator in both hemispheres. It is difficult to measure the rate of migration, but in the southern hemisphere there is an indication that the migration may reach 20° latitude in 2011. In previous cycles, solar maximum has occurred when the migration reached ~ 20°.

This analysis does not indicate the strength of the maximum.

**Acknowledgements.** The observations used herein are the result of a cooperative program of the Air Force Research Laboratory and the National Solar Observatory. I am grateful for the assistance of NSO personnel, especially John Cornett, Timothy Henry, Lou Gilliam, and Wayne Jones, for observing and data-reduction and analysis services and maintenance of the Evans Solar Facility and its instrumentation. Finally, I am extremely grateful to the recently passed Raymond N. Smartt, who completely redesigned the Sacramento Peak Photoelectric Coronal Photometer filters in 1982, making it the excellent instrument it is today. I thank the anonymous referee for his suggestions for improving this paper.